# All-sky homogeneity of precipitable water vapour over Paranal

Richard R. Querel*[a,b], Florian Kerber[c]

[a]National Institute of Water and Atmospheric Research (NIWA), Lauder, New Zealand;
[b]Department of Electrical Engineering, University of Chile, Santiago de Chile, Chile;
[c]European Southern Observatory, Karl-Schwarzschild-Str. 2, 85748 Garching, Germany

## ABSTRACT

A Low Humidity and Temperature Profiling (LHATPRO) microwave radiometer, manufactured by Radiometer Physics GmbH (RPG), is used to monitor sky conditions over ESO's Paranal observatory in support of VLT science operations. The unit measures several channels across the strong water vapour emission line at 183 GHz, necessary for resolving the low levels of precipitable water vapour (PWV) that are prevalent on Paranal (median ~2.4 mm). The instrument consists of a humidity profiler (183-191 GHz), a temperature profiler (51-58 GHz), and an infrared camera (~10 µm) for cloud detection. We present, for the first time, a statistical analysis of the homogeneity of all-sky PWV using 21 months of periodic (every 6 hours) all-sky scans from the radiometer. These data provide unique insight into the spatial and temporal variation of atmospheric conditions relevant for astronomical observations, particularly in the infrared. We find the PWV over Paranal to be remarkably homogeneous across the sky down to 27.5° elevation with a median variation of 0.32 mm (peak to valley) or 0.07 mm (rms). The homogeneity is a function of the absolute PWV but the relative variation is fairly constant at 10-15% (peak to valley) and 3% (rms). Such variations will not be a significant issue for analysis of astronomical data. Users at ESO can specify PWV – measured at zenith – as an ambient constraint in service mode to enable, for instance, very demanding observations in the infrared that can only be conducted during periods of very good atmospheric transmission and hence low PWV. We conclude that in general it will not be necessary to add another observing constraint for PWV homogeneity to ensure integrity of observations. For demanding observations requiring very low PWV, where the relative variation is higher, the optimum support could be provided by observing with the LHATPRO in the same line-of-sight simultaneously. Such a mode of operations has already been tested but will have to be justified in terms of scientific gain before implementation can be considered. This will be explored further in the future.

## 1. INTRODUCTION

Water vapour is the main source of opacity in the Earth's atmosphere at infrared (IR) wavelengths. That opacity is the primary reason that ground-based astronomical observations are only made through certain wavelength regions, called windows, because of their relative lack of water absorption lines and other telluric absorption features. These windows are relatively well-defined and stable for most observations at temperate latitudes and moderate elevations above sea level. Atmospheric transmission is a function of the depth of air above the observer. This is true for almost all atmospheric constituents since gases such as $O_2$ are well-mixed throughout the atmosphere. In contrast, water can coexist in three phases at atmospheric temperatures, and hence, its atmospheric distribution is highly variable in altitude as well as geographical location and time. Atmospheric water vapour content can be expressed as precipitable water vapour (PWV) which is the equivalent condensed amount of water in an atmospheric column above the observer (as a depth in units of mm).

ESO's Very Large Telescope (VLT) is located on Cerro Paranal (24.6°S, 70.4°W, 2635 m.asl). Since November 2011 ESO operates a water vapour monitor in support of VLT science operations[1]. Paranal is a very dry site with a median PWV of ~2.4 mm offering excellent conditions for IR observations when real-time PWV information is taken into account. Its pronounced seasonal variations are well-documented[2]. In contrast the variation of uniformity of PWV across the sky at any given moment has not been studied.

*richard.querel@niwa.co.nz

Progress in astrophysical research depends more than ever on a quantitative analysis of astronomical observations. Hence there is increasing demand by the community for observatories to provide high-quality calibration of the instrumentation that supports such analysis of the data. As a consequence ESO has introduced PWV as a new observational constraint that can be specified by the user for service mode observations (section 1.3). This PWV value is of course given for the zenith and one important question in this context is the intrinsic atmospheric variation of PWV across the sky. The PWV variation over Paranal as recorded by the LHATPRO water vapour monitor is the subject of this contribution. The characteristics for Cerro Armazones, the site of the future European Extremely Large Telescope (E-ELT), about 20 km from Paranal, are very similar[3] and our results will be applicable there as well.

## 1.1 Low Humidity and Temperature Profiling microwave radiometer (LHATPRO)

The Low Humidity And Temperature PROfiling microwave radiometer (LHATPRO), manufactured by Radiometer Physics GmbH (RPG), measures the atmosphere at two frequency ranges focusing on two prominent emission features: a strong $H_2O$ line (183 GHz) and an $O_2$ band (51-58 GHz). Using 6 and 7 channels, respectively, the radiometer retrieves the profile of humidity and temperature up to an altitude of ~12 km[1,4] The spatial resolution is given by the size of the radiometer beam (1.4° FWHM). For a full calibration with absolute standards an additional external calibration target, cooled down to the boiling point of liquid nitrogen (LN2), is used. The LHATPRO offers a measurement duty cycle of >97%. Details of the radiometer are described in Rose et al[4].

The LHATPRO water vapour radiometer (WVR) was commissioned during a 2.5 week period in October and November 2011 during which the calibration was tested with respect to balloon-borne radiosondes, the established standard in atmospheric physics[1]. The LHATPRO measures PWV in the range 0-20 mm with an accuracy of better than 0.1 mm and an internal precision of 30 µm. The 183 GHz line is intrinsically very strong and reliable readings can be obtained even in the driest of conditions encountered on Paranal. In addition, ESO's LHATPRO is equipped with an IR radiometer (10.5 µm) to measure sky brightness temperature down to -100° C. This allows for detection of high altitude clouds, e.g. cirrus, that consist of ice crystals but contain practically no liquid water (see Kerber, Querel and Hanuschik[5] this conference). The IR radiometer can by fully synchronized for scanning with the microwave radiometer.

Before deployment of the LHATPRO efforts to monitor water vapour on Paranal used dedicated standard star observations at the 8-m unit telescopes (UTs) with the spectroscopic instruments UVES, X-Shooter, CRIRES and VISIR. See ESO's instrumentation webpage for details: http://www.eso.org/sci/facilities/paranal/instruments.html

The method for deriving PWV from such spectra taken at various wavelengths by fitting the observed spectrum with an atmospheric radiative transfer model is described in Querel et al[6]. These routine monitoring data are available at the ESO Quality Control website:

http://www.eso.org/observing/dfo/quality/GENERAL/PWV/HEALTH/trend_report_ambient_PWV_closeup_HC.html

For CRIRES and VISIR, dedicated observations of the sky background are taken in wavelength regions with strong $H_2O$ emission features at 5.05 µm and 19.5 µm, respectively. Again these data are automatically processed and fit with an atmospheric model. Results available at: http://www.eso.org/sci/facilities/paranal/sciops/CALISTA/pwv/data/2012.html. The above VLT instruments were validated with respect to radiosondes during 2009 as part of the E-ELT site characterisation campaigns[2] and provide a PWV accuracy of about 20%.

## 1.2 Assessing PWV variations over Paranal

The above approach for measuring PWV is inherently unsuitable for obtaining information on the spatial homogeneity of PWV. With the VLT instruments observations are restricted to a pencil-beam (line-of-sight) and any attempt to sample a significant part of the sky would require prohibitive (expensive) amounts of observing time on an 8-m telescope. In practice, single standard star observations towards an arbitrary – but low air mass – position in the sky are taken and the sampling is very sparse in time (maximum a few per night). The achievable accuracy is limited and hence small variations towards different lines of sight cannot be easily detected with good confidence. No attempt has ever been made to measure the variation of PWV across the sky over Paranal with that approach. The only recorded indications of spatial variations are folded into time-series measurements of the same standard star during E-ELT site testing campaigns[2].

The LHATPRO radiometer is far superior for assessing PWV variations for a number of reasons. The radiometer provides continuous high time resolution (1 s cadence) measurements of PWV with high precision and accuracy – a factor of 10 better than the spectroscopic methods[6]. Most importantly the LHAPRO has an all-sky pointing capability that can be used to conduct user-defined scans of the sky in an automated fashion. The WVR can be commanded to repeat an observation sequence n-times resulting in months of unsupervised operations if needed. The IR channel is mounted on the same azimuth stage and can be instructed to perform its elevation scan in lock with the water vapour radiometer resulting in fully synchronous and parallel observations in the IR channel as well. The operational scheme currently set-up on Paranal and covering a period of 24 hours is the following:

- 2-D all-sky scan: step size of 12° azimuth, 12.5° elevation (duration 6 min, repeated every 6 h; 1.7% of day)
- Cone scan at 30° elevation (Hovmöller): step size 6° (duration 2.5 min, repeated every 15 min; 16.7% of day)
- Routine observations set to zenith-staring mode (81.6% of day)

All data can be corrected for air mass and so PWV at zenith is available nearly continuously and is displayed in the Paranal VLT control room.

The scanning sequence for the two-dimensional all-sky scan is as follows: the observation begins from its routine state staring at zenith (elevation = 90°, azimuth = 0°). The mirror tips down in elevation towards the horizon in incremental steps of 12.5°. This continues until elevation = 27.5°. The mirror is capable of scanning to below the horizon, but for these sky-scans the limit has been set to 27.5° to minimize reflections from nearby buildings and objects in the line-of-sight. Once at this lower elevation limit, the radiometer rotates in azimuth +12° and then repeats the elevation steps back up to 77.5°, then moves a further +12° in azimuth and continues to zig-zag in this way across the sky until the azimuth mount has completed one full revolution. It is important to note that there is no overlap in this scheme and that the PWV at zenith is measured only once at the beginning of the 6 minute scan. The resulting data are processed by the RPG software and displayed in real-time (Figure 1).

The 2-D scans acquired in this manner and taken during the period June 2012 to March 2014 (21 months) form the basis for the analysis presented in sections 2 and 3.

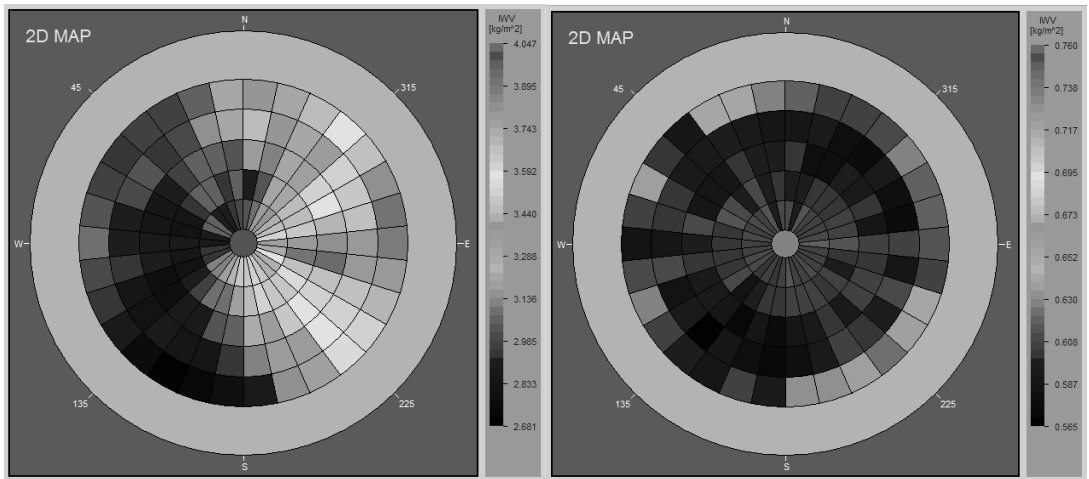

Figure 1. Examples of 2-dimensional sky scans (down to 27.5° elevation) as they are produced by the LHATPRO radiometer on Paranal every 6 hours. These products are a direct output of the RPG software and are displayed in real-time in the VLT control room. (Left): The sky on August 17, 2012 at 16:08 UT which was characterized by a significant gradient and variation of PWV. (Right): The relatively homogeneous and dry sky measured July 21, 2012 at 08:31 UT.

**1.3 Science operations and user-supplied atmospheric constraints**

The VLT on Paranal hosts a variety of IR instruments: with VISIR[7] and MIDI[8] covering the mid-IR domain which is most affected by variations in PWV. In support of IR observations a water vapour radiometer was deployed on Paranal in 2011 as part of the VISIR upgrade project[9].

ESO has been successfully using the concept of service mode observations for many years now. There are two elements to it: a) users can provide observing constraints required for meeting their scientific goals; and, b) a queue-based operational model allows the support astronomers at the observatory to flexibly react to changes in ambient conditions in order to optimally satisfy the requested constraints.

This is most important for rare conditions that are very unlikely to be encountered during a short visit at the observatory. In the past, an IR instrument like VISIR has almost exclusively been used during "bright time" (moonlit conditions) regardless of the atmospheric properties relevant for the IR simply because no independent means of accessing PWV had been available. Now that a stand-alone PWV monitor is available, PWV (at zenith) has been introduced as a user-provided observing constraint, just like astronomical seeing or lunar phase. In service mode, scientific programs requiring very demanding constraints are given highest priority if the relevant constraint(s) is (are) met.

One key question in this context is the intrinsic variation of PWV across the sky. The timing of observations of astronomical targets in service mode is driven by conditions and practical constraints at the observatory. Hence the pointing in the sky is not pre-determined although air mass constraints can be given. If the intrinsic PWV variation is small, the PWV at zenith will be a good indicator for arbitrary pointings after air mass correction. If the PWV variations are large, a second constraint limiting the permitted variation may be required to ensure the integrity of the science observations. For example, photometric observations of faint targets requiring long integration times could be compromised by variations in the transmission caused by PWV variations. Note that "photometric" conditions are given here by sky transmission (PWV), even in the absence of any clouds. Hence the quantification of intrinsic PWV variations across the sky is relevant for science operations.

Conditions on Paranal (median PWV 2.4 mm) can be very dry. Recently (July 5, 2012), a "dry episode" with PWV ~0.1 mm lasting for 12 hours has been reported by Kerber et al[10]. This was related to the excursion of Antarctic air to the Atacama Desert. During these spectacular conditions the transparency of the atmosphere will increase even outside the established atmospheric windows. A case in point is the Paschen alpha line at 1875 nm. This region is completely opaque in median conditions on Paranal but offers a transmission of about 40% at 0.5 mm[10]. Clearly, significant variations of PWV could limit the feasibility of such observations overall. Thus one specific aspect of the work presented here is to explore whether variations of PWV across the sky at very low absolute PWV can be expected to be an issue for such observations.

## 2. DATA ANALYSIS

Following an evaluation campaign on the Zugspitze in southern Germany, the LHATPRO was installed on Paranal in October 2011. Some changes to its settings and retrieval software were performed during this commissioning period that left some artifacts in the processed data. In this study we have included measurements made from July 1, 2012 through March 31, 2014 (21 months). During this period there were 2588 all-sky scans performed.

The instrument measures the line-of-sight water vapour path and reports it as PWV. This value is proportional to the air mass of the given observation path. Using the homogeneous plane-parallel atmosphere approximation for air mass: air mass = secant(zenith angle); all PWV values were corrected for air mass, resulting in zenith path values at each measurement point across the all-sky scan. All comparisons in this study were performed using air mass corrected values relative to the actual zenith water vapour path measured at zenith.

In order to remove known artificial spikes in the data set (i.e. direct sunlight, reflections off nearby telescope enclosures, etc.), the mean and standard deviation for each scan was calculated, and air mass corrected data elements that were greater than five sigma from the zenith PWV value were removed. Following this step, the mean and standard deviation were recalculated using the filtered data set.

# 3. RESULTS

For illustration we provide all-sky plots of representative PWV conditions over Paranal. For low PWV, median and high PWV, respectively examples of homogeneous and variable PWV are given. This is followed by a statistical analysis.

## 3.1 Low PWV conditions

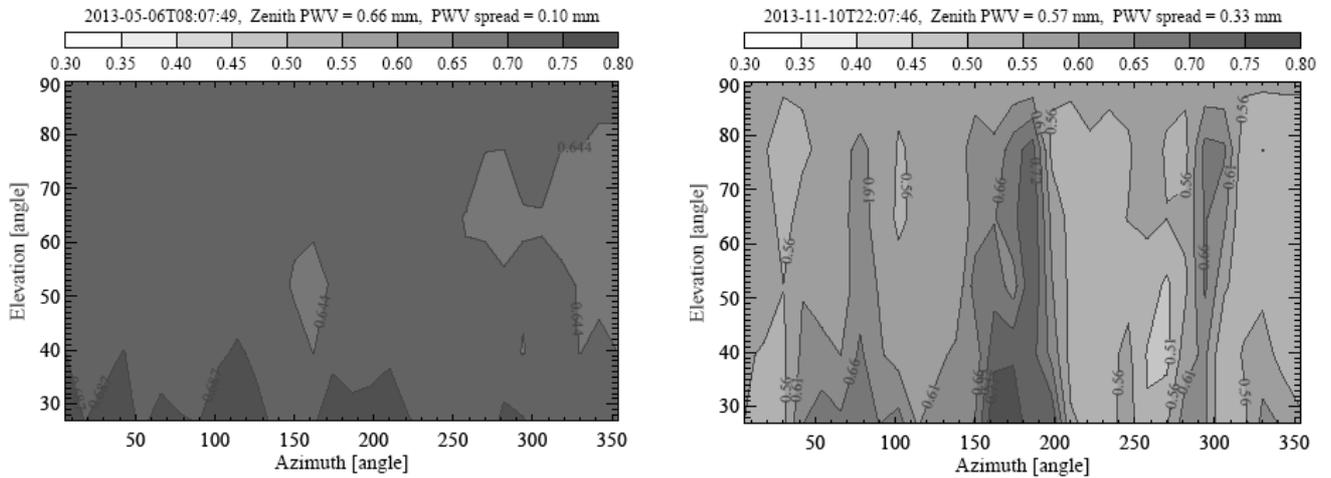

Figure 2. Azimuth-Elevation plot showing the distribution of PWV across the sky down to 27.5° elevation. Left: Example of low PWV conditions (0.7 mm) with small variations. Right: Example of low PWV conditions (0.6 mm) with large variations.

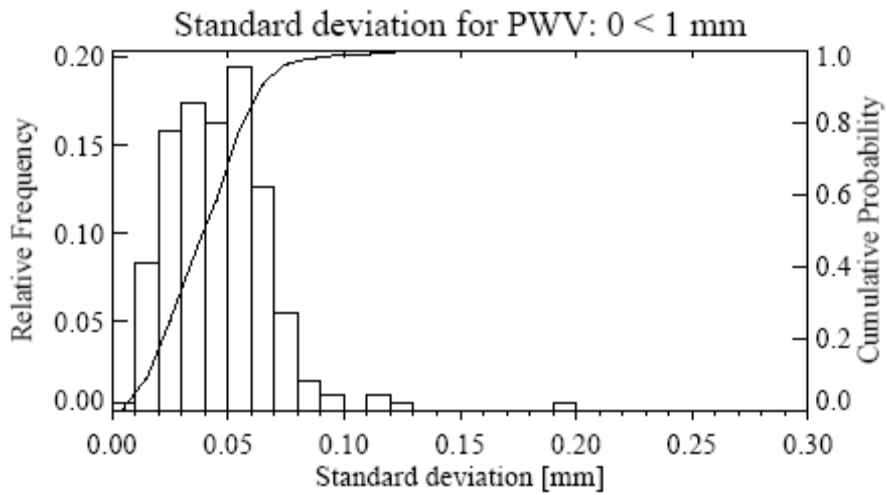

Figure 3. Histogram showing the distribution of PWV across the sky down to 27.5° elevation for low PWV < 1 mm conditions

## 3.2 Median PWV conditions

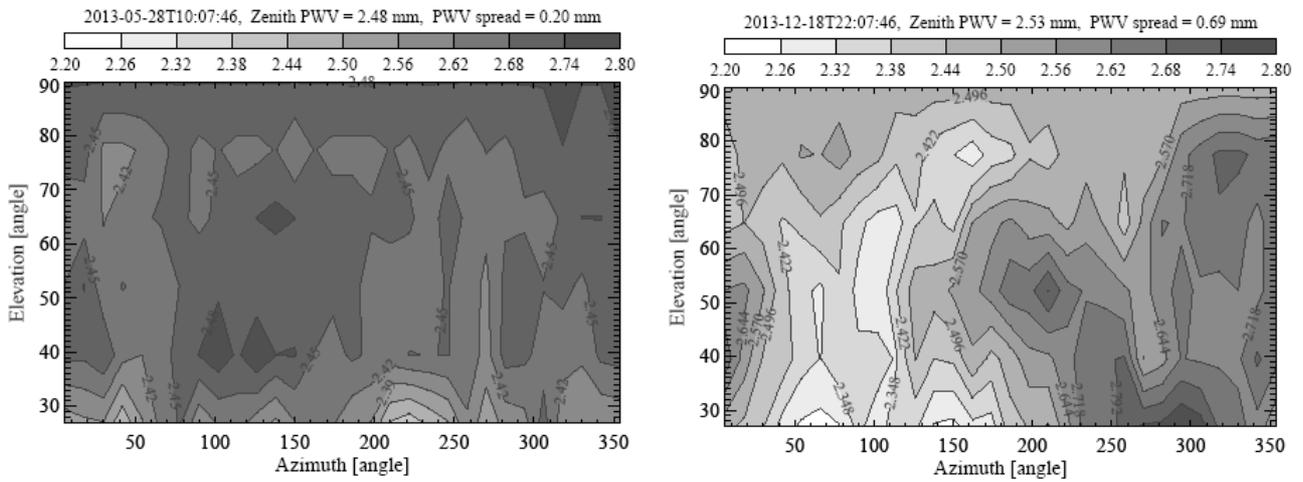

Figure 4. Azimuth-Elevation plot showing the distribution of PWV across the sky down to 27.5° elevation. Left: Example of median PWV conditions (2.5 mm) with small variations. Right: Example of median PWV conditions (2.5 mm) with large variations.

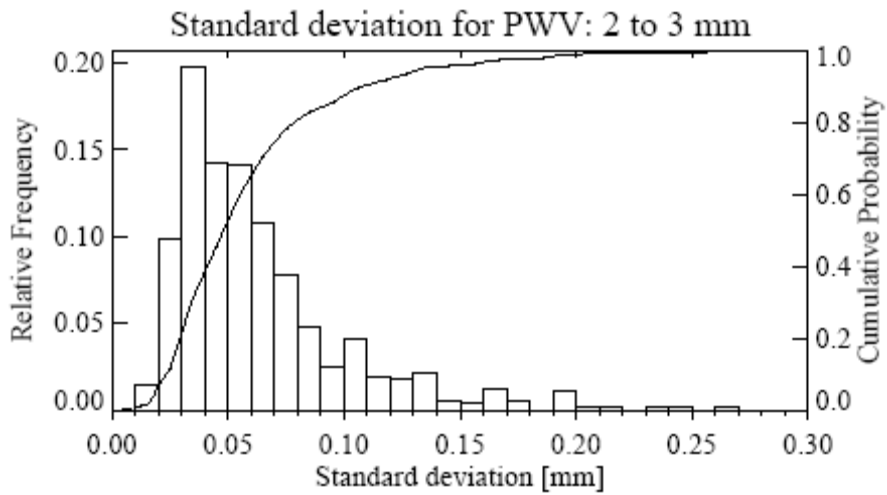

Figure 5. Histogram showing the distribution of PWV across the sky down to 27.5° elevation for median (2.4 mm) conditions; bin used is 2 mm < PWV < 3 mm.

## 3.3 High PWV conditions

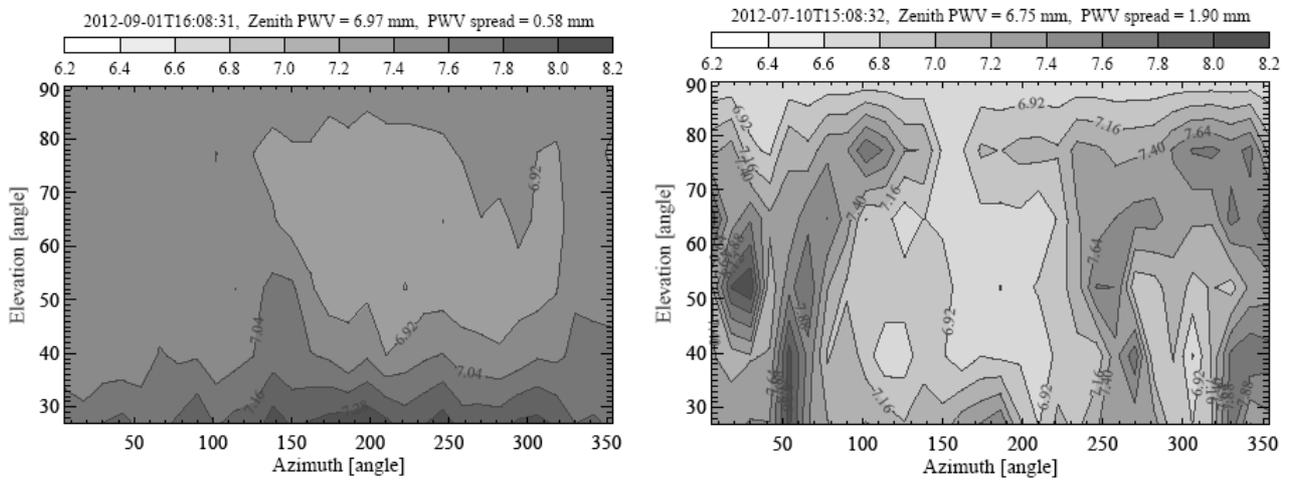

Figure 6. Azimuth-Elevation plot showing the distribution of PWV across the sky down to 27.5° elevation. Left: Example of high PWV conditions (7.0 mm) with small variations. Right: Example of high PWV conditions (6.8 mm) with large variations.

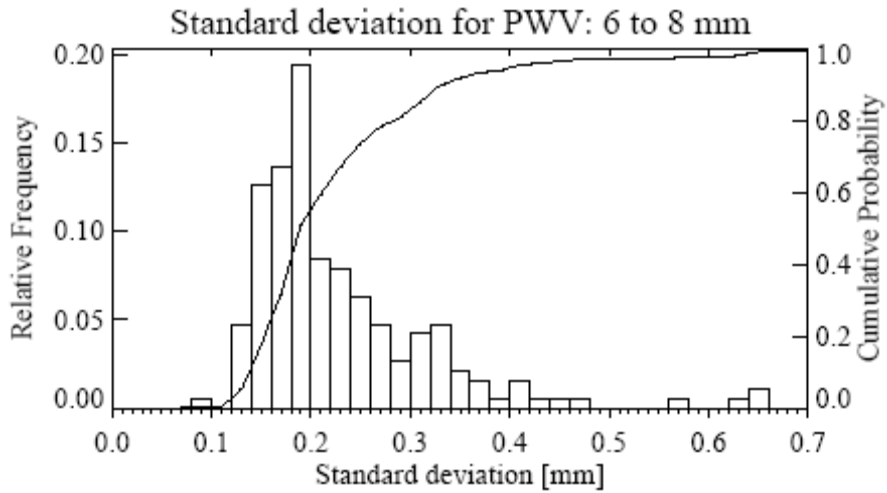

Figure 7. Histogram showing the distribution of PWV across the sky down to 27.5° elevation for high 6 mm < PWV < 8 mm conditions.

## 3.4 Statistical Analysis of sky conditions over Paranal

For an observatory and its users it is very important to know what atmospheric conditions can be expected for the scientific observations. For service mode in particular, in which many astronomers make use of setting constraints in order to optimize the scientific return of their observations, statistical information on the availability of certain parameters such as PWV is relevant and of large practical value. Here we present the first such analysis for the homogeneity of PWV across the sky over Paranal.

Table 1. Homogeneity of PWV over Paranal. Shown in the table is the variation in PWV across the sky as: Standard deviation [mm] and [%]; Measured spread in PWV, peak to valley (PtV), [mm] and [%]. Using 2588 sky-scans over 21 months, representing day and night conditions, the 10, 25, 50, 75, and 90 percentiles were computed. The median values are shown in boldface.

| Percentiles | PWV variation SDev [mm] | PWV variation SDev [%] | PWV Variation PtV [mm] | PWV Variation PtV [%] |
|---|---|---|---|---|
| 10 | 0.03 | 1.5 | 0.17 | 8 |
| 25 | 0.05 | 2.1 | 0.22 | 10 |
| **50** | **0.07** | **3.1** | **0.32** | **15** |
| 75 | 0.13 | 4.7 | 0.64 | 22 |
| 90 | 0.22 | 6.4 | 1.04 | 30 |

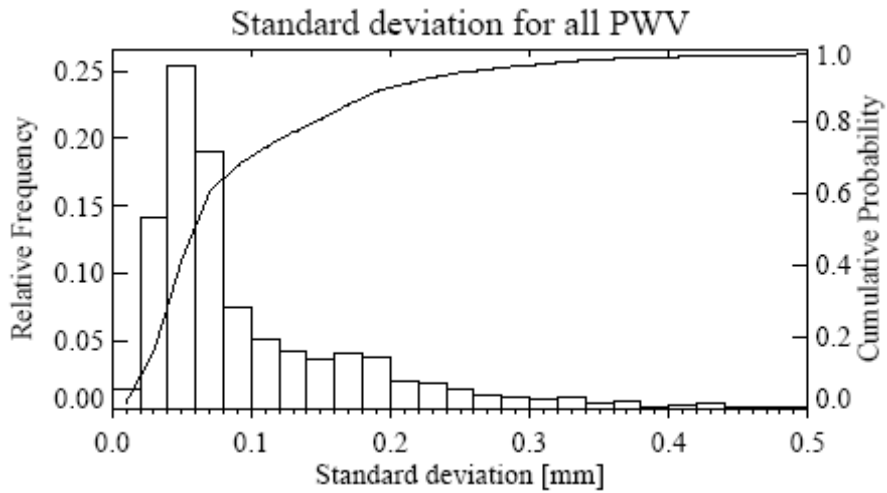

Figure 8. Histogram and cumulative probability distribution of the PWV variation of across the sky down to 27.5° elevation.

Table 2. The standard deviation [mm] of the PWV variations binned by zenith PWV value. Using 2588 sky-scans over 21 months, representing day and night conditions, the 10, 25, 50, 75, and 90 percentiles were computed. The median values are shown in boldface.

| Percentiles | 0-1 mm | 1-2 mm | 2-3 mm | 3-4 mm | 4-6 mm | 6-8 mm | 8-10 mm | >10mm |
|---|---|---|---|---|---|---|---|---|
| 10 | 0.02 | 0.04 | 0.03 | 0.04 | 0.07 | 0.15 | 0.14 | 0.11 |
| 25 | 0.03 | 0.05 | 0.04 | 0.06 | 0.09 | 0.17 | 0.16 | 0.15 |
| **50** | **0.05** | **0.06** | **0.05** | **0.08** | **0.14** | **0.20** | **0.20** | **0.26** |
| 75 | 0.06 | 0.07 | 0.08 | 0.12 | 0.20 | 0.27 | 0.26 | 0.48 |
| 90 | 0.07 | 0.09 | 0.11 | 0.18 | 0.26 | 0.35 | 0.33 | 0.89 |
| Count | 252 | 818 | 567 | 262 | 332 | 191 | 92 | 104 |

Table 3. The relative standard deviation [%] of the PWV variations binned by zenith PWV value. Using 2588 sky-scans over 21 months, representing day and night conditions, the 10, 25, 50, 75, and 90 percentiles were computed. The median values are shown in boldface.

| Percentiles | 0-1 mm | 1-2 mm | 2-3 mm | 3-4 mm | 4-6 mm | 6-8 mm | 8-10 mm | >10mm |
|---|---|---|---|---|---|---|---|---|
| 10 | 3.7 | 2.1 | 1.3 | 1.3 | 1.5 | 2.2 | 1.5 | 1.1 |
| 25 | 4.8 | 2.9 | 1.5 | 1.7 | 2.0 | 2.5 | 1.8 | 1.4 |
| **50** | **6.0** | **4.0** | **2.2** | **2.3** | **2.8** | **2.9** | **2.2** | **2.3** |
| 75 | 7.8 | 5.2 | 3.1 | 3.5 | 4.0 | 3.9 | 3.0 | 3.9 |
| 90 | 12.9 | 6.4 | 4.7 | 5.4 | 5.1 | 5.1 | 3.9 | 5.8 |
| Count | 252 | 818 | 567 | 262 | 332 | 191 | 92 | 104 |

Table 4. The measured spread in PWV [mm], peak to valley, of the PWV variations binned by zenith PWV value. Using 2588 sky-scans over 21 months, representing day and night conditions, the 10, 25, 50, 75, and 90 percentiles were computed. The median values are shown in boldface.

| Percentiles | 0-1 mm | 1-2 mm | 2-3 mm | 3-4 mm | 4-6 mm | 6-8 mm | 8-10 mm | >10mm |
|---|---|---|---|---|---|---|---|---|
| 10 | 0.11 | 0.17 | 0.15 | 0.24 | 0.37 | 0.65 | 0.59 | 0.54 |
| 25 | 0.16 | 0.21 | 0.19 | 0.30 | 0.46 | 0.76 | 0.70 | 0.72 |
| **50** | **0.20** | **0.25** | **0.27** | **0.41** | **0.68** | **0.91** | **0.90** | **1.09** |
| 75 | 0.25 | 0.31 | 0.39 | 0.63 | 0.99 | 1.19 | 1.19 | 1.78 |
| 90 | 0.33 | 0.40 | 0.60 | 1.00 | 1.34 | 1.61 | 1.61 | 3.04 |
| Count | 252 | 818 | 567 | 262 | 332 | 191 | 92 | 104 |

Table 5. The relative measured spread in PWV [%], peak to valley, of the PWV variations binned by zenith PWV value. Using 2588 sky-scans over 21 months, representing day and night conditions, the 10, 25, 50, 75, and 90 percentiles were computed. The median values are shown in boldface.

| Percentiles | 0-1 mm | 1-2 mm | 2-3 mm | 3-4 mm | 4-6 mm | 6-8 mm | 8-10 mm | >10mm |
|---|---|---|---|---|---|---|---|---|
| 10 | 19 | 10 | 7 | 7 | 8 | 9 | 7 | 5 |
| 25 | 22 | 13 | 8 | 9 | 10 | 11 | 8 | 6 |
| **50** | **26** | **17** | **11** | **12** | **14** | **13** | **10** | **10** |
| 75 | 37 | 22 | 16 | 18 | 20 | 19 | 14 | 15 |
| 90 | 61 | 28 | 24 | 28 | 27 | 24 | 18 | 20 |
| Count | 252 | 818 | 567 | 262 | 332 | 191 | 92 | 104 |

## 4. OUTLOOK

The uniformity of PWV across the sky over Paranal is remarkably high. This suggests that a PWV constraint for zenith will be sufficient to secure the correct conditions for arbitrary pointings for most science applications. As described in section 1.3 PWV can be very low on Paranal enabling new science taking advantage of increased atmospheric transmission outside of established astronomical windows. This kind of science is directly dependent on the actual transmission and hence PWV along the line-of-sight towards the target. Since the variation of PWV is, relatively speaking, the highest for low PWV conditions such variations may in fact negatively impact photometric measurements. Therefore it may be prudent to consider introducing an additional constraint on the PWV representing its variation during an observation. A more in-depth analysis of the all-sky PWV data set, including diurnal and seasonal trends in the variations is in preparation.

Even better support to the science exploitation of such data could be given by measuring the PWV along the same air column traversed by the science light from the target (line-of-sight) simultaneously with the science observation. The LHATPRO WVR is capable of all-sky pointing and tracking and thus line-of-sight support. The spatial resolution is limited by the size of the radiometer beam (1.4° FWHM). This mode has been technically implemented by RPG and has been tested on Paranal. Full implementation of such a mode will require careful planning and would have to be justified by clear scientific gain. In any case it would be limited to particularly demanding programmes of low PWV science.

In parallel, progress in atmospheric modeling using input from the LHATPRO has been demonstrated to be a valid alternative to the standard observations of telluric stars. This is a field of active development at ESO.

## 5. SUMMARY

We have analysed the homogeneity of PWV over ESO's Paranal observatory using data from a dedicated WVR LHATPRO. Using all-sky scans (down to 27.5° elevation) that are routinely taken every 6 h and which cover a period of 21 months we find the PWV over Paranal to be remarkably uniform with a median variation of 0.32 mm (peak to valley) or 0.07 mm (rms). The homogeneity is a function of the absolute PWV but the relative variation is fairly constant at 10% (peak to valley) and 3% (rms). Such variations will not be a significant issue for analysis of astronomical data. These results will be used to provide guidance for the astronomers using ESO's service mode. Optimal support for demanding low PWV science could be given by providing simultaneous line-of-sight observations with LHATPRO pointed towards the science target.

**Acknowledgements:** RQ acknowledges funding from Conicyt through Fondecyt grant 3120150.

**Product Disclaimer** Certain commercial equipment, instruments, or materials are identified in this report in order to specify the experimental procedure adequately. Such identification is not intended to imply recommendation or endorsement by the European Southern Observatory, nor is it intended to imply that the materials or equipment identified are necessarily the best available for the purpose.